\documentclass[proceedings]{JHEP}
\usepackage{epsfig}
\conference{TMR meeting, Paris, 1999}
\title{Stable non-BPS D-branes of type I}
\author{M. Frau$^1$, L. Gallot$^1$, A. Lerda$^{2,1}$ and P. Strigazzi$^1$ \\
$^1$ Dipartimento di Fisica Teorica, Universit\`a di Torino\\
Via P. Giuria 1, I-10125 Torino, Italy\\
$^2$ Dipartimento di Scienze e Tecnologie Avanzate\\
Universit\`a del Piemonte Orientale, I-15100 Alessandria, Italy}
\abstract{We review the boundary state description of the non-BPS 
D-branes in the type I string theory and show that the only stable 
configurations are the D-particle and the D-instanton.
We also compute the gauge and gravitational interactions of the non-BPS 
D-particles and compare them with the interactions of the 
dual non-BPS particles of the heterotic string 
finding complete agreement. In this way we provide further dynamical evidence 
of the heterotic/type I duality. }
\newcommand{\bra}[1]{\langle{#1}|}
\newcommand{\ket}[1]{|{#1}\rangle}

\def\beq{\begin{equation}}
\def\eeq{\end{equation}}
\def\beqa{\begin{eqnarray}}
\def\eeqa{\end{eqnarray}}

\newcommand{\EQ}{\begin{equation}}
\newcommand{\EN}{\end{equation}}
\newcommand{\bea}{\begin{eqnarray}}
\newcommand{\ena}{\end{eqnarray}}

\renewcommand{\a}{\alpha}

\newcommand{\shalf}{\frac{1}{2}}

\renewcommand{\thefootnote}{\fnsymbol{footnote}}
\def\one{{\hbox{ 1\kern-.8mm l}}}

\def\NS{{\rm NS}}

\def\ii{{\rm i}}
\def\ee{{\rm e}}

\keywords{D-branes, Boundary states}
\begin{document}
\renewcommand{\thefootnote}{\arabic{footnote}}
\setcounter{footnote}{0}
\setcounter{page}{1}
Dirichlet branes (or D-branes for short)
are a key ingredient in our understanding of the duality 
relations between superstring theories. They
are described by a boundary 
conformal field theory, and admit a two-fold interpretation: 
on the one hand, 
the D-branes are objects on which open 
strings can end, and on the other hand they  
can emit or absorb closed strings \cite{Pol}. These two 
descriptions can be related to each other by world-sheet duality. 
 
Introducing D-branes in a theory of closed strings amounts 
to extend their conformal field theory by introducing
world-sheets with boundaries and imposing appropriate boundary
conditions on the closed string coordinates $X^\mu$. In the operator
formalism these boundary conditions are implemented through
the so called boun-dary state \begin{footnote}{For a recent review on 
the boundary state formalism and its applications, 
see Ref.~\cite{pa}}\end{footnote} $\ket{Dp}$,
whose bosonic part is defined by the following eigenvalue 
problem
\begin{eqnarray}
\partial_{\tau}X^{\alpha}(\sigma , 0)\, \ket{Dp}_X &=& 0~~,
\nonumber\\
\Big(X^{i}(\sigma , 0)-x^i\Big)\, \ket{Dp}_X &=& 0~~,
\label{boundcond} 
\end{eqnarray}
where the index $\alpha=0,\ldots,p$ labels the longitudinal
directions, the index $i=p+1,\ldots,9$ labels the transverse
directions and the $x^i$'s denote the position of the brane in the 
transverse space.
{\it World-sheet} supersymmetry requires that analogous equations must
be also imposed on the left and right moving fermionic fields 
$\psi^\mu$ and ${\widetilde \psi}^\mu$. These equations, which define 
the fermionic part of the boundary state, are
\begin{eqnarray}
\Big(\psi^\alpha(\sigma,0) - {\rm i}\,\eta\,{\widetilde \psi}^\alpha(\sigma,0)
\Big)\,\ket{Dp,\eta}_\psi = 0 ~~,
\nonumber \\
\Big(\psi^i(\sigma,0) +{\rm i}\,\eta\,{\widetilde \psi}^i(\sigma,0)
\Big)\,\ket{Dp,\eta}_\psi = 0 ~~,
\label{boundcond1} 
\end{eqnarray}
where $\eta=\pm1$. Notice that there are two consistent implementations
of the fermionic  boundary conditions corresponding to the
sign of $\eta$, and consequently there are two different boundary states
\begin{equation}
\ket{Dp,\eta} = \ket{Dp}_X\,\ket{Dp,\eta}_\psi
\label{bs}
\end{equation}
both in the NS-NS and in the R-R sectors.
The overlap equations (\ref{boundcond}) and (\ref{boundcond1})
allow to determine the explicit structure of the boundary
states (\ref{bs}) up to an overall factor. This normalization
can then be uniquely fixed by factorizing amplitudes with 
closed strings emitted from a disk 
\cite{many,cpb} and turns out
to be given by (one half of) the brane tension measured in
units of the gravitational coupling constant, {\it i.e.} 
\begin{equation}
T_p=\sqrt{\pi}\,\Big(2\pi\sqrt{\alpha'}\Big)^{3-p}~~.
\nonumber
\end{equation}
We would like to remark that even if each boundary state
$\ket{Dp,\eta}$ is perfectly consistent from the conformal
field theory point of view, not all of them are
acceptable in string theory. In fact, to describe a 
physical D-brane a boundary state has to
satisfy three requirements \cite{BerGab}: 
\begin{itemize}
\item[{\it i})] to be invariant under the closed string GSO projection (and 
also under orbifold or orientifold projections if needed); 
\item [{\it ii})] the tree level amplitude due to the exchange of 
closed strings between two boundary sta-tes, after modular transformation in 
the open string channel, has to make sense as a consistent open string 
partition function at one-loop;
\item[{\it iii})] the open strings introduced through the D-branes must have 
consistent couplings with the original closed 
strings~\footnote{This last condition, which is rather difficult to prove 
in general, does not give more constraints than the first two in the case 
of type I and II theories.}.
\end{itemize}

Using these prescriptions, it is rather simple
to find the boundary state for the supersymmetric BPS D$p$-branes 
of type II.
In particular, the GSO projection of the type II theories 
forces us to retain only the following linear combinations
\begin{eqnarray}
\ket{Dp}_{\rm NS} &=& 
\frac{1}{2} \Big[ \ket{Dp,+}_{\rm NS} 
- \ket{Dp,-}_{\rm NS} \Big] \nonumber \\
\ket{Dp}_{\rm R} &=& 
\frac{1}{2} \Big[ \ket{Dp,+}_{\rm R} 
+\ket{Dp,-}_{\rm R} \Big]
\label{bsgso}
\end{eqnarray}   
in the NS-NS and in the R-R sectors respectively, with $p=0,2,4,6,8$ for IIA, 
$p=-1,1,3,5,7,9$ for IIB. 
The normalization of the boundary states (\ref{bsgso})
can be deduced by requiring that
the spectrum of the open strings living on the D$p$-brane 
(called $p$-$p$ strings)
be supersymmetric.
To read the spectrum of these open strings from the boundary state, 
one has first to evaluate 
the closed string exchange amplitude
\begin{equation}
\bra{Dp}\,P\,\ket{Dp} ~~,
\nonumber
\end{equation}
where $P$ is the closed string propagator
\begin{equation}
P= \frac{\alpha'}{2}\,\int_0^\infty\!
 dt~\,{\rm e}^{-t(L_0+{\widetilde L}_0 -2a)}
\nonumber 
\end{equation}
(with $a_{\rm NS}=1/2$ and $a_{\rm R}=0$), and then perform the modular
transformation $t\to 1/s$ to exhibit the open string
channel. Applying this procedure, one finds 
the following relations
\begin{eqnarray}
{}_{\rm NS}\bra{Dp,\eta}\,P\,\ket{Dp,\eta}_{\rm NS} 
&=& \int_0^\infty\frac{ds}{s}\,{\rm Tr}_{\rm NS} \,q^{2L_0-1}~~,\nonumber \\
{}_{\rm NS}\bra{Dp,\eta}\,P\,\ket{Dp,-\eta}_{\rm NS} &=&
\int_0^\infty\frac{ds}{s}\,{\rm Tr}_{\rm R}\, q^{2L_0} ~~,\nonumber \\
{}_{\rm R}\bra{Dp,\eta}\,P\,\ket{Dp,\eta}_{\rm R} \nonumber \\
=\int_0^\infty\frac{ds}{s}&&\!\!\!\!\!\!\!\!{\rm Tr}_{\rm NS}\,(-1)^F\,
 q^{2L_0-1}~~,
\label{openclosed} \\
{}_{\rm R}\bra{Dp,\eta}\,P\,\ket{Dp,-\eta}_{\rm R} \nonumber \\ 
=\int_0^\infty\frac{ds}{s}&&\!\!\!\!\!\!\!\!{\rm Tr}_{\rm R}\,(-1)^F\,
 q^{2L_0} = 0~~,
\nonumber
\end{eqnarray} 
where $q={\rm e}^{-\pi s}$. It is then clear that in order to
obtain the
supersymmetric ({\it i.e.} GSO projected) open string amplitude 
\begin{eqnarray}
\int_0^\infty\frac{ds}{s}\left[{\rm Tr}_{\rm NS}\left(\frac{1+(-1)^F}{2}
\right)q^{2L_0-1} \right.\nonumber \\
-\left.{\rm Tr}_{\rm R}\left(\frac{1+(-1)^F}{2}
\right)q^{2L_0} \right]
\end{eqnarray}
one must consider the following boundary state
\begin{equation}
\ket{Dp} = \ket{Dp}_{\rm NS} \pm \ket{Dp}_{\rm R}
\label{bsbps}
\end{equation}
where the sign ambiguity is related to the existence of branes and anti-branes.
Note that both the NS-NS and the R-R components of the boundary state
(\ref{bsbps})
have the same normalization so that the tension of 
a D$p$-brane essentially equals the density of its charge under the R-R
potential: this is the BPS relation which is typical
of the supersymmetric and stable branes of type II. 

The criteria $i)$ - $iii)$
defining physical D-bra-nes do not rely 
at all on {\it space-time} supersymmetry, and thus one may wonder whether 
in type II theories there may exist also non-supersymmetric branes. 
This problem has been systematically addressed in a series of
papers by A. Sen \cite{sen1} - \cite{sen4}, who constructed 
explicit examples of non-BPS (and hence non-supersymmetric) branes.
In particular in Ref.~\cite{sen3}, he considered the superposition of a 
D-string of type IIB and an anti-D-string (with a $Z_2$ Wilson line on it) 
and by suitably 
condensing the tachyons of the open strings stretching between the brane
and the anti-brane, he managed to construct a new configuration of type IIB 
which behaves like a D-particle, does not couple to any R-R field 
and is heavier by a factor of $\sqrt{2}$ than the
BPS D-particle of the IIA theory. The boundary state for this non-BPS 
D-brane has been explicitlely constructed in Ref.~\cite{gallot}.
This 
construction can be obviously generalized to the case 
of a pair formed by two BPS D$(p+1)$-branes with opposite R-R charge
(and with a $Z_2$ Wilson line) which, after tachyon condensation, 
becomes a non-BPS D$p$-brane. 
Alternatively, this same non-BPS
configuration can be described starting from a
superposition of two BPS D$p$ branes with opposite R-R charge 
and modding out the theory by the operator $(-1)^{F_L}$ whose effect is to 
change the sign of all states in the R-R and R-NS sectors. In this second
scheme, a superposition of a D$p$-brane and anti-D$p$-brane of type IIA (IIB)
becomes in the reduced theory a non-BPS D$p$-brane of type IIB (IIA).
In either way we therefore find that there
exist non-BPS D$p$-branes for $p=0,2,4,6,8$. For reviews on this subject,
see Refs.~\cite{reviews,SCHWARZ}.

These branes are manifestly 
non-supersym-metric, but nevertheless they satisfy 
the conditions ${\it i})$ - ${\it iii})$ mentioned above, and thus
are perfectly consistent from the closed string point of view.
In particular, since they are not charged under any R-R field, 
the boundary state for these non-BPS D-branes has
only the NS-NS component, namely
\begin{equation}
\ket{Dp} = 
\mu_p \,
\ket{Dp}_{\rm NS}
\label{tens1}
\end{equation} 
where we have introduced a (positive) 
coefficient $\mu_p$ to allow for a different
normalization with respect to the standard BPS case. This normalization
can be deduced by requiring that the closed string amplitude 
between two non-BPS branes, after a modular transformation,
has the interpretation of the partition 
function of a non-supersymmetric ({\it i.e.} without the GSO 
projection~\begin{footnote}{Note that the non-supersymmetric GSO projection 
$(1-(-1)^{F})/2$ cannot correspond to a {\it single} brane since 
all NS open string zero-modes are projected out.}\end{footnote} )
open string model. Indeed, by requiring that
\begin{equation}
\bra{Dp} P \ket{Dp} =
\int_0^\infty\frac{ds}{s}\left[{\rm Tr}_{\rm NS}
\,q^{2L_0-1} 
-{\rm Tr}_{\rm R}\,q^{2L_0} \right]
\label{ns}
\end{equation}
we find that $\mu_p=\sqrt{2}$, thus confirming that the non-BPS 
D-branes are
heavier by a factor of $\sqrt{2}$ than the corresponding BPS ones. 
Although the-se non-BPS D-branes of type II may 
have interesting properties \cite{Harvey}, 
it is clear from (\ref{ns}) that they are not stable,
because the absence of the GSO projection on the open strings
leaves the NS tachyon on their world-volume. 
However, these non-BPS branes could become stable
in an orbifold of the type II theory, say IIA(B)$/{\cal P}$, 
provided that the tachyon be odd under the projection ${\cal P}$. 
In the orbifold theory, the non-BPS 
vacuum amplitude of the $p$-$p$ open-strings 
is clearly given by
\begin{eqnarray}
{\cal Z}_{open}=n
\int_0^\infty\frac{ds}{s}\left[{\rm Tr}_{\rm NS}\left(\frac{1+{\cal P}}{2}
\,q^{2L_0-1} \right)\right.\nonumber \\
-\left.{\rm Tr}_{\rm R}\left(\frac{1+{\cal P}}{2}
\,q^{2L_0}\right) \right]
\label{zop}
\end{eqnarray}
where $n$ is a positive integer representing some possible multiplicity. 
(In our present discussion we take $n=1$ for the sake of simplicity, 
but the case $n=2$ will appear later.)  The natural question to 
ask now is to which boundary state the amplitude (\ref{zop})
could correspond.
In the case of a space-time orbifold, the perturbative spectrum 
of the bulk theory contains only 
closed strings which can be untwisted (U) or twisted (T) under the 
orbifold. Therefore, there are four 
sectors to which the bosonic states belong, namely
(NS-NS;U), (R-R;U), (NS-NS;T) or (R-R;T), and the-re exist different types
of boundary states depending on which components in those sectors
they have. For example, when the orbifold projection ${\cal P}$ acts 
as the inversion of some space-time coordinates, 
the boundary state which gives rise to 
(\ref{zop}) turns out \cite{sen2,BerGab,GabSen}
to have only a component in the unstwisted NS-NS
sector and another in the twisted R-R sector, {\it i.e.} 
\begin{equation}
\ket{Dp} = {1 \over \sqrt{2}} \left( \sqrt{2}\ket{Dp}_{\rm NS;U}+ \sqrt{2}
\ket{Dp}_{\rm R;T} \right)~~.
\label{bsut}
\end{equation}
In particular, if one computes the exchange amplitude 
$\bra{Dp}\,P\,\ket{Dp}$ with
this boundary state and performs a modular transformation,
one can see that the terms of (\ref{zop})
with ${\cal P}$ originate precisely 
from the twisted part of the boundary state.

In the case of a world-sheet orbifold, however, this simple 
picture does not hold. To illustrate this point, we
consider the specific case of the type I theory
which is the orbifold of the IIB theory by the world-sheet parity $\Omega$
\cite{ps}. The distinctive feature of this model is that 
the perturbative states of the
twisted sector of the {\it bulk} theory now 
correspond to unoriented {\it open} 
strings which should then be appropriately incorporated in the
boundary state formalism. Let us briefly summarize how
this is done (for more details see \cite{PCAI}).
The starting point is the projection of the closed string spectrum 
onto states which are invariant under $\Omega$. 
The corresponding closed string partition function is
obtained by adding a Klein bottle 
contribution to the modular invariant (halved) torus contribution. 
The Klein bottle is a genus one non-orientable 
self-intersecting surface which may be seen equivalently
as a cylinder ending at two crosscaps. A crosscap is a line of 
non-orientability, a circle with opposite points identified, and thus the 
associated crosscap state $\ket{C}$ is defined by
\begin{eqnarray}
X^{\mu}(\sigma+\pi , 0)\,\ket{C}&=& X^{\mu}(\sigma , 0)\,\ket{C} ~~,\\
\partial_{\tau}X^{\mu}(\sigma+\pi , 0)\,\ket{C} &=& -\partial_{\tau}X^{\mu}
(\sigma , 0)\,\ket{C}~~, \nonumber
\label{crosscap}
\end{eqnarray}
and by the analogous relations appropriate for world-sheet fermions. 
As is clear from these equations, the crosscap state 
does not have any space-time interpretation but nevertheless it is 
related to the boundary state of the BPS space-time filling D9 brane
through
\begin{equation}
\ket{C} ~ \propto ~{\rm i}^{L_0+\tilde{L}_0}\,\ket {D9}~~.
\label{cd9}
\end{equation} 
The normalization of $\ket{C}$,
which may be fixed up to an overall sign using
the action of $\Omega$ on the massless closed string modes and 
the world-sheet duality, turns out to be
32 times the normalization of the boundary state for the D9-brane. 
Consequently, the (negative) charge 
for the unphysical 10-form R-R potential created by the
crosscap must be compensated by
the introduction of 32 D9 branes. 
In this way we then introduce unoriented open strings 
which start and end on these 32 D9 branes, whose
vacuum amplitude is given by
\begin{eqnarray}
{\cal Z}_{open} &=&
{1 \over 2} \Big(2^{10} \bra{D9}P\ket{D9}
\label{ZopI}  \\
&&+\,\,2^5 \bra{D9}P\ket{C} + 2^5\bra{C}P\ket{D9}\Big)~~,
\nonumber
\end{eqnarray}
where the first line represents the contribution of the
annulus and the second line the contribution of the
M\"obius strip.
By adding to (\ref{ZopI}) 
the contribution of the Klein bottle we obtain 
a modular invariant expression, in which the tadpoles 
for the massless unphysical states cancel  
if and only if we choose the still unfixed 
overall sign in front of the crosscap state to be +. A 
moment thought shows that this corresponds 
to choose the open string gauge group to be $SO(32)$ 
(the other sign instead leads to the gauge group $Sp(32)$). 
Thus, we can say that the type I theory possesses a 
``background'' boundary state given by
\begin{equation}
{1 \over \sqrt{2}} \Big(\ket{C}+32\ket{D9} \Big)~~.
\label{background}
\end{equation}
where the factor of $1/\sqrt{2}$ has been introduced 
to obtain the right normalization of the various spectra.
Performing a modular transformation, we can rewrite
the amplitude ${\cal Z}_{open}$ of eq. (\ref{ZopI})
in the open string channel
as follows
\begin{eqnarray}
\int_0^\infty\frac{ds}{s}
\left[{\rm Tr}_{\rm NS}\left( {{1+(-1)^{F} \over 2}} 
{1+ \Omega \over 2 }\,q^{2L_0-1}\right)\right.\nonumber \\
\left.-\,\,{\rm Tr}_{\rm R}\left( {{1+(-1)^{F} \over 2}} 
{1+ \Omega \over 2 }\,q^{2L_0}\right)\right]~~,
\nonumber
\end{eqnarray}
where the part depending on $\Omega$ comes from the M\"obius contribution. 
Thus, we see that in the type I theory the crosscap state plays 
the same role that the twisted part of the boundary state had 
in the space-time orbifolds. 
In the following, we shall use this remark in order to 
classify the stable non-BPS branes of type I theory.

We have seen before that the type IIB theory contains unstable
non-BPS D$p$-branes with $p=0,2,4,6,8$ which are described
by the boundary state (\ref{tens1}). Now, we address
the question whether these D-branes become stable in the
type I theory, {\it i.e.} we examine whether the tachyons
of the $p$-$p$ open strings are removed by $\Omega$. As
explained in \cite{gallot}, the world-sheet parity can be
used to project the spectrum of the $p$-$p$ strings
only if $p=0,4,8$. Thus, the non-BPS D2 and D6 branes
will not be further considered. However, in order to be 
exhaustive, we must take into account also another
kind of configuration, namely the superposition of
a D$p$-brane and an anti-D$p$-brane of type IIB.
This pair clearly does not carry any R-R charge,
is represented by a boundary state of the form
(\ref{tens1}) and is unstable due to the presence
of tachyons in the open strings stretching between the
brane and the anti-brane. In the type I theory, however,
these tachyons might be projected out. A systematic
analysis \cite{WITTEN,gallot} shows that in this case
$\Omega$ can be used as a projection only if $p=-1,3,7$. 

In conclusion, we have to analyze the stability of the
non-BPS D$p$-branes of type I with $p=-1,0,3,4,7,8$
whose corresponding boundary states $\ket{Dp}$
are given by eq. (\ref{tens1}) with suitable values of
$\mu_p$.
To address this problem, we need to consider the spectrum
of the unoriented strings living on the
brane world-volume (the $p$-$p$ sector),
and also the spectrum of the open strings
stretched between the D$p$-brane
and each one of the 32 D9-branes of the background
(the $p$-$9 \oplus 9$-$p$ sector), in which tachyonic
modes could develop.

Let us first analyze the $p$-$p$ sector, whose
total vacuum amplitude is given by
\begin{equation} 
{\cal A}_{\rm tot}=\frac{1}{2}\Big({\cal A}+{\cal M}+{\cal M}^\ast\Big)
\label{atot}
\end{equation}
where ${\cal A}$ and ${\cal M}$ are respectively the
annulus and the M\"obius strip contributions
\begin{equation}
{\cal A} = \bra{Dp} P \ket{Dp} \quad \mbox{and}
\quad
{\cal M} = \bra{Dp} P \ket{C}~~.
\label{amplo}
\end{equation}
After a modular transformation, in the
open string channel these amplitudes read respectively
\begin{eqnarray}
{\cal A} &=& \mu_p^2\, V_{p+1} \,(8 \pi^2 \a')^{- {p+1 \over 2}} 
\int_0^{\infty}  {ds \over 2s} \, s^{- {p+1 \over 2}} 
\nonumber \\
&\times& \left[ \,{ f_3^8(q) - f_2^8 (q) \over f_1^8(q)} \,\right] ~~,
\label{anulus}
\end{eqnarray}
and 
\begin{eqnarray}
{\cal M} &=&  2^{7-p \over 2} \, \mu_p \, V_{p+1} \,
 (8 \pi^2 \a')^{- {p+1 \over 2}} 
\int_0^{\infty} {ds \over 2s} \, s^{- {p+1 \over 2}}
\nonumber \\
&&\left[\, {\rm e}^{\ii (p-9) \pi/4}\, {f_4^{p-1}(\ii\,q)\,f_3^{9-p}(\ii\,q)
 \over f_1^{p-1}(\ii\,q)\,f_2^{9-p}(\ii\,q) } \right.  
\nonumber \\
&&\left.- {\rm e}^{\ii (9-p) \pi/4}\,{f_3^{p-1} (\ii\,q)\, f_4^{9-p}(\ii\,q)
 \over f_1^{p-1} (\ii\,q)\,f_2^{9-p}(\ii\,q)}\, \right]~~,
\label{moebius}
\end{eqnarray}
where $f_1$, $f_2$ $f_3$ and $f_4$ are 
the standard one-loop functions defined for example in Ref.
\cite{Pol}.
The spectrum of the $p$-$p$
open strings can be analyzed by expanding the total
amplitude ${\cal A}_{\rm tot}$
in powers of $q$. The leading term in this expansion is
\begin{eqnarray}
{\cal A}_{\rm tot} &\sim& \int_0^{\infty} {ds \over 2s}\, 
s^{- {p+1 \over 2}}\,q^{-1}\nonumber\\
&&\times\,\,\Big[ \mu_p^2 - 2 \,\mu_p \sin \big( {\pi \over 4}(9-p)\big) \Big] 
~~.
\label{asymptotic}
\end{eqnarray}
The $q^{-1}$ behavior of the integrand signals the presence
of tachyons in the spectrum; 
therefore, in order not to have them, we must 
require that 
\begin{equation}
\mu_p = 2\,\sin \big(\frac{\pi}{4}(9-p)\big)~~.
\label{stability}
\end{equation}
Since $\mu_p$ has to be positive,
the only possible solutions are
\begin{equation}
 \begin{array}{|c|c|c|c|c|}\hline
 p     & -1 &    0     & \,7\, &      8        \\ \hline
 \mu_p &  2 & \sqrt{2} & 2 & \sqrt{2}      \\ \hline
 \end{array}
\label{tabular}
\end{equation}
{F}rom this table we see that in the type I theory there exist two
even non-BPS but stable D$p$-branes: 
the D-particle and the D8-brane. 
Both of them have a tension that is a factor
of $\sqrt{2}$ bigger than the corresponding BPS branes of the
type IIA theory.
Moreover, there exist two 
odd non-BPS but stable D$p$-branes of type I: 
the D-instanton and the D7-brane.
Their tension is twice the one of the corresponding
type IIB branes, in accordance with the fact that, as mentioned above, 
they can be simply interpreted as the superposition of a brane
with an anti brane, so that the R-R part of the boundary state
cancels while the NS-NS part doubles.

This classification of the stable non-BPS D-branes of type I based
on the table (\ref{tabular}) 
is in complete agreement with the results of Refs.~\cite{WITTEN,HG}
derived from the K-theory of space-time.

Let us now analyze the $p$-$9 \oplus 9$-$p$ sector. The
relevant quantity to consider is the "mixed" cylinder
amplitude
\begin{equation}
{\cal A}_{\rm mix}
={32 \over 2} \Big(\bra{Dp}P\ket{D9} +\bra{D9}P\ket{Dp}\Big)~~,
\label{p9}
\end{equation}
which, after a modular transformation in the open string channel,
reads
\begin{eqnarray}
{\cal A}_{\rm mix}&=& 2^5\, \mu_p\, V_{p+1}\, (8 \pi^2 \a')^{- {p+1 \over 2}}
\, \int_0^{\infty}  {ds \over 2s} \,s^{- {p+1 \over 2}} \nonumber \\
&\times& \left[ { f_3^{p-1}(q)f_2^{9-p} (q) \over f_1^{p-1}(q)f_4^{9-p}(q)}-
 { f_2^{p-1}(q)f_3^{9-p} (q) \over f_1^{p-1}(q)f_4^{9-p}(q)} \right]~~,
\nonumber
\end{eqnarray} 
where the first and second term in the square brackets
account respectively for the 
NS and R sector. This expression needs some comments. First, for $p=-1,0$
we see that there are no tachyons in the spectrum; moreover,
the values of $\mu_p$ for the D-instanton and D-particle 
are crucial in order to 
obtain a sensible partition function for open strings stretching between the 
non-BPS objects and the 32 D9-branes. Indeed, they are the smallest 
ones that make integer the coefficients in the partition functions.
Secondly, for $p=7,8$ we directly
see the existence of a NS tachyon,  so 
that the corresponding branes are actually unstable \cite{gallot}. 
Hence, only the D-instanton and the D-particle are
fully stable configuration of type I string theory \cite{gallot,SCHWARZ}. 
Nevertheless, the strict relation connecting the D0-brane and the D(-1)-brane 
to the D8-brane and the D7-brane respectively suggests however that also the
latter may 
have some non trivial meaning.
Finally, we observe that the zero-modes of the Ramond sector
of these $p$-$9 \oplus 9$-$p$ strings
are responsible for the degeneracy of the non-BPS D$p$-branes
under the gauge group SO(32): in particular the
D-particle has the
degeneracy of the spinor representation of SO(32), as
discussed in \cite{sen3,sen4}. Thus the D-particle accounts 
for the existence in type I of the non-perturbative non-BPS states 
required by the heterotic/type I duality.

These same methods may be used in order to study the stability of a non-BPS 
D$p$-brane in presence of another D$q$-brane. Indeed the spectrum 
of open strings stretching between two such (distant) objects at rest has a 
vacuum amplitude given by
\begin{equation}
{\mu_p\mu_q\over 2} \Big({}_{\NS}\bra{Dp}P\ket{Dq}_{\NS}+
{}_{\NS}\bra{Dq}P\ket{Dp}_{\NS}\Big)~~.
\label{pq}
\end{equation} 
The overall factor of one-half 
indicates that, respectively to the IIB case, only the $\Omega$ symmetric 
combinations are retained. By explicitly computing
this amplitude, one can see that for $|p-q|\leq 3$ and for
sufficiently small values of the distance between the branes,
a NS tachyon develops in the open string spectrum, thus
signalling the unstability of the configuration. As a first consequence of 
this, we can conclude that the superposition of two non-BPS D-particles 
with trivial quantum numbers, decays into the vacuum \cite{sen3}. 
As a matter of fact, a stable non-BPS D$p$-brane 
is its own anti-brane, as may be also inferred from the K-theory analysis 
which shows that the conserved D-brane charge in that case is $Z_2$ valued
\cite{WITTEN}.
This analysis shows that there is no hope to form a stable superposition of 
$N$ non-BPS D$p$-branes of type I as they always exert an attractive 
force on each other, as may be seen from (\ref{pq}). This is to be 
contrasted with the case of a space-time orbifold \cite{GabSen} where, 
for some particular values of the compactification radii, a compensation 
occurs between the attractive force due to exchange of untwisted NS-NS 
states and the repulsive force due to exchange of twisted R-R states.   
As a second consequence, and for analogous reasons, the superposition of a 
D-particle and of a D1-string is unstable and decays in the vacuum. 
Note that in that case, the 0-1 open string is T-dual of the 8-9 string 
responsible for the unstability of the D8-brane.  

Up to now we have investigated the flat ten dimensional case.
However, this analysis can be easily extended also to the case
in which some directions are compactified. Contrarily 
to the BPS D-branes, the non-BPS one are not stable in 
all moduli space. As an example, let us consider the non-BPS
D-particle and compactify one space direction along a circle of radius $R$.
Then, one can observe that 
a tachyon develops in the $0$-$0$ open string sector if $R<R_c=
\sqrt{\alpha'/2}$,
so that below the critical radius $R_c$ the configuration 
is unstable. The corresponding stable non-BPS configuration which 
carries the same quantum numbers in this range of moduli is a superposition 
of wrapped D1 and anti-D1 strings with a $Z_2$ Wilson line. Notice that when 
the time direction is compactified, the D-particle is stable for any 
value of the radius.

\bigskip 

We now present the basic
ideas and results about the gravitational
and gauge interactions of two stable non-BPS
D-particles of type I string theory (the detailed calculations and analysis
of these interactions can be found in \cite{gls}).

In the type I theory, D-branes interact 
via exchanges of both closed and open bulk strings. Since the dominant 
diagram for open strings has the topology of a disk, 
it gives a subleading (in the string coupling constant) contribution to 
the diffusion amplitude of two branes 
which is thus dominated by the cylinder diagram, {\it i.e.} by 
the exchange of closed strings. In the long distance limit, this
accounts for the gravitational interactions.
Let us now use this observation to calculate the 
dominant part of the scattering 
amplitude between two D-particles of type I moving with a relative
velocity $v$. This process can be simply analyzed 
using the boundary state formalism as explained in \cite{CANGEMI}.
What we need to compute is the cylinder 
amplitude between the boundary state of a  static D-particle $\ket{D0}$
and the boundary state of a moving D-particle $\ket{D0,v}$. The latter
is simply obtained by acting with a Lorentz boost on $\ket{D0}$, {\it i.e.} 
\begin{equation}
|D0,v \rangle = \ee^{\ii \pi \theta J_{01}}\ket{D0}~~,
\end{equation} 
where $\theta$ is the rapidity along direction of motion (which we have
taken to be $x^1$) defined through 
$v={\rm tanh}\pi\theta$, and $J_{01}$ is the corresponding Lorentz generator. 
Thus, the amplitude we are looking for is
\begin{equation}
{{\cal A}} = \langle D0 | P | D0,v \rangle +
\langle D0,v | P | D0 \rangle~~,
\label{amplitude}
\end{equation}
which indeed reduces to (\ref{pq}) for $v\to 0$.
From this expression, we can extract the long range gravitational 
potential energy, which, in the non relativistic limit,
reads \cite{gls}
\begin{eqnarray}
V^{\rm grav}(r) =  
(2 \kappa_{10})^2 \,{M_0^2 \over 7\, \Omega_8\, r^7}
\nonumber \\
\times \left(1 + \shalf \,v^2 +o(v^2) \, \right)~~, 
\label{newton1}
\eeqa
where $r$ is the radial coordinate, $\Omega_8$ is the 
area of the unit $8$-dimensional sphere,
$M_0=T_0/\kappa_{10}$ is the D-particle mass and $\kappa_{10}$
is the gravitational coupling constant in ten dimensions. 
Hence the boundary state calculation correctly reproduces
the gravitational potential we expect for a pair of D-particles 
in relative motion.

Although they are subdominant in the string coupling constant, 
the interactions of the D-parti-cle with the open strings of the bulk
are nevertheless interesting because they account for the gauge interactions.
Since the non-BPS D-particles of type I are spinors of $SO(32)$, 
their gauge coupling 
is fixed by the spinorial representation 
they carry (except possibly by the overall strength).
The stringy description of such a coupling
has been provided in \cite{gls} where we have
shown that it is represented by an open string diagram with the 
topology of a disk with two boundary components, one lying
on the D9-branes from which the gauge boson is 
emitted, and the other lying on the D-particle (see Figure 1). 

\EPSFIGURE[ht]{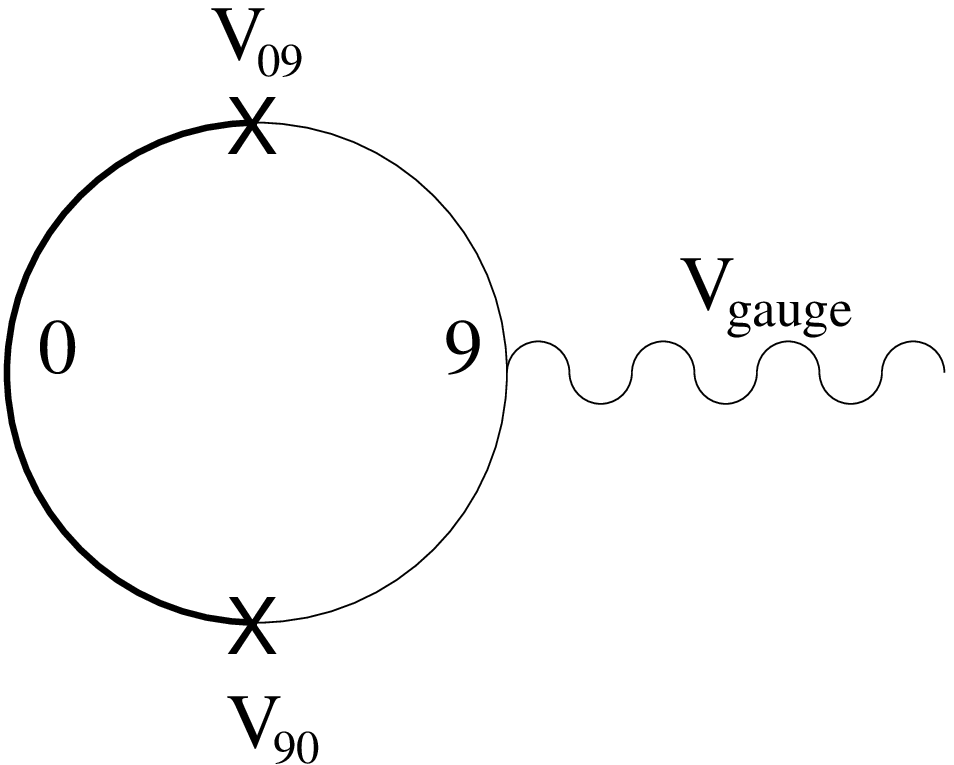,height=4cm,width=5cm}
{The disk diagram describing the gauge coupling of a type I D-particle.}

At the points where the two boundary components join, we thus
have to insert a vertex operator $V_{90}$ (or $V_{09}$) that
induces the transition from Neumann to 
Dirichlet (or from Dirichlet to Neumann) boundary conditions in the nine 
space directions. 
As we have mentioned before, the $SO(32)$ 
degeneracy of the D-particle is due to the fermionic massless modes of the
open strings stretching between the D-particle and each of the 32 D9-branes;
therefore it is natural to think that
the boundary changing operators $V_{90}$
and $V_{09}$ are given by the vertex operators for these
massless fermionic modes
\cite{gls}. By construction, these operators carry Chan Paton factors
in the fundamental representation of SO(32), while the vertex 
operator $V_{\rm gauge}$ for the gauge boson carries a
Chan-Paton factor in the adjoint. 
As a consequence, the diagram represented in Figure 1 must be considered as 
the one point function of the gauge boson in the background formed by a 
D-particle seen as an object in the bi-fundamental representation
of $SO(32)$. Hence, we do 
not see the entire gauge degeneracy of the D-particle because the 
degrees of freedom we use to describe it are not accurate enough. This is 
reminiscent from the fact that, in the boundary state formalism, also
the Lorentz 
degeneracy of a D-brane is hidden. 
Using this result, we can easily compute the 
Coulomb potential energy $V^{\rm gauge}(r)$ for two
D-particles placed at a distance $r$. Indeed, this is
simply obtained by gluing two diagrams like
that of Figure 1
with a gauge boson propagator, and its explicit expression
turns out to be \cite{gls}
\beqa
{V}^{\rm gauge}(r) =
-\, \frac{g_{\rm YM}^2}{2}~\frac{1}{7\,\Omega_8\,r^7} \nonumber \\
\times ~\left(\delta^{AB}\,\delta^{CD}
-\delta^{AC}\,\delta^{DB}\right)~~,
\label{vgauge0}
\eeqa
where $g_{\rm YM}$ is the gauge coupling constant in
ten dimensional type I string theory, and $A$, $B$, $C$ and  $D$
are indices in the fundamental representation of $SO(32)$.

We conclude by recalling that the non-BPS D-particles of type I are dual to
perturbative non-BPS states of the $SO(32)$ heterotic string which also 
have gravitational and gauge interactions among themselves. These can
be computed using standard perturbative methods and if one takes into
account the known duality relations and renormalization effects, one can 
explicitly check that they agree with the expressions (\ref{newton1})
and (\ref{vgauge0}). This agreement provides further dynamical
evidence of the heterotic/type I duality beyond the BPS level. 

\end{document}